\documentclass[10pt,A4paper]{article}
\usepackage{amsmath}
\usepackage{amsfonts}
\usepackage{graphicx}
\usepackage{graphics}
\usepackage{theorem}
\usepackage{color}

\usepackage{amsxtra}
\usepackage{amstext}
\usepackage{amssymb}
\usepackage{latexsym}

\usepackage[square]{natbib}
\usepackage{graphicx}
\usepackage{dcolumn}
\usepackage{bm}

\makeatletter
\newcommand{\cn}{\mathop{\operator@font cn}}
\makeatother
\makeatletter
\newcommand{\sn}{\mathop{\operator@font sn}}
\makeatother
\makeatletter
\newcommand{\dn}{\mathop{\operator@font dn}}
\makeatother

\newcommand{\GA}{GALI}
\newcommand{\Ly}{Lyapunov exponents}

\theoremstyle{plain}
\newtheorem{theorem}{Theorem}
\begin{document}

\nocite{*}
\bibliographystyle{plain}

\title{Detecting Order and Chaos by the Linear Dependence Index (LDI) Method}
\author{Chris ANTONOPOULOS\footnote{E-mail: {\tt antonop@math.upatras.gr}},
Tassos BOUNTIS\footnote{E-mail: {\tt bountis@math.upatras.gr}}}
\date{} \maketitle
\begin{center}
{\it Department of Mathematics\\and\\Center for Research and
Applications of Nonlinear Systems (CRANS),\\University of Patras,\\
GR--$26500$, Rio, Patras, Greece}
\end{center}

\begin{abstract}
We introduce a new methodology for a fast and reliable
discrimination between ordered and chaotic orbits in
multidimensional Hamiltonian systems which we call the Linear
Dependence Index (LDI). The new method is based on the recently
introduced theory of the Generalized Alignment Indices (GALI). LDI
takes advantage of the linear dependence (or independence) of
deviation vectors of a given orbit, using the method of Singular
Value Decomposition at every time step of the integration. We show
that the LDI produces estimates which numerically coincide with
those of the GALI method for the same number of $m$ deviation
vectors, while its main advantage is that it requires considerable
less CPU time than GALI especially in Hamiltonian systems of many
degrees of freedom.
\end{abstract}

\section{Introduction}

The fast and reliable discrimination of chaotic and ordered orbits
of conservative dynamical systems is of crucial interest in many
problems of nonlinear science. By the term conservative we
characterize here systems which preserve phase space volume (or some
other positive function of the phase space variables) during time
evolution. Important examples in this class are $N$ -- degree of
freedom (dof) Hamiltonian systems and $2N$ -- dimensional symplectic
maps. As is well known, in such systems chaotic and regular orbits
are distributed in phase space in very complicated ways, which often
makes it very difficult to distinguish between them. In recent
years, several methods have been developed and applied to various
problems of physical interest in an effort to distinguish
efficiently between ordered and chaotic dynamics. Their
discrimination abilities and overall performance, however, varies
significantly, making some of them more preferable than others in
certain situations.

One of the most common approaches is to extract information about
the nature of a given orbit from the dynamics of small deviations,
evaluating the maximal Lyapunov Characteristic Exponent (LCE)
$\sigma_1$. If $\sigma_1 > 0$ the orbit is characterized as chaotic.
The theory of Lyapunov exponents was first applied to characterize
chaotic orbits by Oseledec \cite{O68}, while the connection between
Lyapunov exponents and exponential divergence of nearby orbits was
given in \cite{BGS76,P77}. Benettin et al. \cite{BGGS80a} studied
the problem of the computation of all LCEs theoretically and
proposed in \cite{BGGS80b} an algorithm for their efficient
numerical computation. In particular, $\sigma_1$ is computed as the
limit for $t \rightarrow \infty$ of the quantity:
\begin{equation}
L_1(t)=\frac{1}{t}\, \ln  \frac{\|\vec{w}(t)\|}{\|\vec{w}(0)\|}\,
,\, \mbox{i.e.}\,\, \sigma_1 = \lim_{t\rightarrow \infty} L_1 (t) \,
, \label{eq:lyap1_def}
\end{equation}
where $\vec{w}(0)$, $\vec{w}(t)$ are deviation vectors from a given
orbit, at times $t=0$ and $t>0$ respectively. It has been shown that
the above limit is finite, independent of the choice of the metric
for the phase space and converges to $\sigma_1$ for almost all
initial vectors $\vec{w}(0)$ \cite{O68,BGGS80a,BGGS80b}. Similarly,
all other LCEs, $\sigma_2$, $\sigma_3$ etc. are computed as limits
for $t \rightarrow \infty$ of some appropriate quantities, $L_2(t)$,
$L_3(t)$ etc. (see for example \cite{BGGS80b}). We note here that
throughout the paper, whenever we need to compute the values of the
maximal LCE or of several LCEs we apply respectively the algorithms
proposed by Benettin et al. \cite{BGS76,BGGS80b}.

Over the years, several variants of this approach have been
introduced to distinguish between order and chaos such as: The Fast
Lyapunov Indicator (FLI) \cite{FLG97,FGL97,FLFF02,GLF02,B05}, the
Mean Exponential Growth of Nearby Orbits (MEGNO) \cite{CS00,CGS03},
the Smaller Alignment Index (SALI)
\cite{S01,Skokos_et._al_PTPS,Skokos_et._al_JPA}, the Relative
Lyapunov Indicator (RLI) \cite{SESF04}, as well as methods based on
the study of power spectra of deviation vectors \cite{VVT00} and
spectra of quantities related to these vectors
\cite{FFL93,LFD93,VC94}.

Recently, the SALI method was generalized to yield a much more
comprehensive approach to study chaos and order in $2N$ --
dimensional conservative systems, called the GALI$_m$ indices
\cite{Skokos_et._al_GALI,AThesis}. These indices represent the
volume elements formed by $m$ deviation vectors ($2\leq m \leq 2N$)
about any reference orbit and have been shown to: (a) Distinguish
the regular or chaotic nature of the orbit faster than other
methods, (b) identify the dimensionality of the space of regular
motion and (c) predict the slow (chaotic) diffusion of orbits, long
before it is observed in the actual oscillations.

In the present paper, we improve the GALI method by introducing the
Linear Dependence Indices (LDI$_m$). The new indices retain the
advantages of the GALI$_m$ and display the same values as GALI, in
regular as well as chaotic cases. More importantly, however, the
computation of the LDI$_m$ is much faster in CPU time, especially if
the dimensionality of phase space becomes large ($N\gg10$). The main
purpose of this paper, therefore, is to strongly advocate the use of
LDI, for the most rapid and efficient study of the dynamics of multi
-- dimensional conservative systems.

For the computation of the LDI$_m$ we use information from the
evolution of $m\geq 2$ deviation vectors from the reference orbit,
as GALI does. However, while GALI requires the computation of many
$m\times m$ determinants at every time step
\cite{Skokos_et._al_GALI,AThesis} in order to evaluate the norm of
the corresponding wedge product, LDI achieves the same purpose
simply by applying Singular Value Decomposition (SVD) to the
$2N\times m$ matrix formed by the deviation vectors. LDI is then
computed as the product of the corresponding singular values of the
above matrix. This not only provides the same numerical values as
the corresponding GALI$_m$, it also requires much less CPU time.

The paper is organized as follows: In section \ref{definition_LDI}
we introduce the new index, explain in detail its computation and
justify its validity theoretically. In section \ref{FPU_example}, we
demonstrate the usefulness of the LDI method, by applying it to the
famous Fermi -- Pasta -- Ulam (FPU) lattice model of $N$ dof, for
small and large $N$. Finally, in section \ref{conclusions} we
present our conclusions, highlighting especially the advantages of
the new index.


\section{Definition of the Linear Dependence Index (LDI)}\label{definition_LDI}
Let us consider the $2N$ -- dimensional phase space of a Hamiltonian
system:
\begin{equation}\label{1}
H\equiv H(q_1(t),\ldots,q_N(t),p_1(t),\ldots,p_N(t))=E
\end{equation}
where $q_i(t),\;i=1,\ldots,N$ are the canonical coordinates,
$p_i(t),\;i=1,\ldots,N$ are the corresponding conjugate momenta and
$E$ is the total energy. The time evolution of an orbit $\vec{x}(t)$
of (\ref{1}) associated with the initial condition:
$$\vec{x}(t_{0})=(q_1(t_{0}),\ldots,q_N(t_{0}),p_1(t_{0}),\ldots,p_N(t_{0}))$$ at initial time $t_0$ is
defined as the solution of the system of $2N$ first order
differential equations (ODE):
\begin{equation}
\frac{dq_i(t)}{dt}=\frac{\partial H}{\partial
p_i(t)},\quad\frac{dp_i(t)}{dt}=-\frac{\partial H}{\partial
q_i(t)},\;i=1,\ldots,N.\label{2}
\end{equation}
Eqs. (\ref{2}) are known as Hamilton's equations of motion and the
reference orbit under study is the solution $\vec{x}(t)$ which
passes by the initial condition $\vec{x}(t_{0})$.

In order to define the Linear Dependence Index (LDI) we need to
introduce the variational equations. These are the corresponding
linearized equations of the ODE (\ref{2}), about the reference orbit
$\vec{x}(t)$ defined by the relation:
\begin{equation}\label{3}
\frac{d\vec{\upsilon_i}(t)}{dt}=\mathcal{J}(\vec{x}(t))\cdot\vec{\upsilon_i}(t),\;i=1,\ldots,2N
\end{equation}
where $\mathcal{J}(\vec{x}(t))$ is the Jacobian of the right hand
side of the system of ODEs (\ref{2}) calculated about the orbit
$\vec{x}(t)$. Vectors
$\vec{\upsilon_i}(t)=(\upsilon_{i,1}(t),\ldots,\upsilon_{i,2N}(t)),\;i=1,\ldots,2N$
are known as deviation vectors and belong to the tangent space of
the reference orbit at every time $t$.

We then choose $m\in[2,2N]$ initially linearly independent deviation
vectors $\vec{\upsilon_m}(0)$ and integrate equation (\ref{3})
together with the equations of motion (\ref{2}). These vectors form
the columns of a $2N\times m$ matrix $\mathcal{A}(t)$ and are taken
to lie along the orthogonal axes of a unit ball in the tangent space
of the orbit $\vec{x}(t)$ so that $\vec{\upsilon}_m(0)$ are
orthonormal. Thus, at every time step, we check the linear
dependence of the deviation vectors by performing Singular Value
Decomposition on $\mathcal{A}(t)$ decomposing it as follows:
\begin{eqnarray}\label{SVD_based_theorem}
\mathcal{A}(t)=U(t)\cdot W(t)\cdot V(t)^{\top},
\end{eqnarray}
where $U(t)$ is a $2N\times m$ matrix, $V(t)$ is an $m\times m$
matrix whose columns are the $\vec{\upsilon}_m(t)$ deviation vectors
and $W(t)$ is a diagonal $m\times m$ matrix, whose entries
$w_1(t),\ldots,w_m(t)$ are zero or positive real numbers. They are
called the {\it singular values} of $\mathcal{A}(t)$. Matrices
$U(t)$ and $V(t)$ are orthogonal so that $U^{\top}(t)\cdot
U(t)=V^{\top}(t)\cdot V(t)=I$, where $I$ is the rectangular
$2N\times 2N$ unit matrix.

We next define the generalized Linear Dependence Index of order $m$
or $\mathrm{LDI}_m$ as the function:
\begin{eqnarray}\label{LDI_definition}
\mathrm{LDI}_m(t)=\prod_{j=1}^m w_j(t)
\end{eqnarray}
with $m=2,3,\ldots,2N$, where $N$ is the number of dof of (\ref{1}).

The reason for defining $\mathrm{LDI}$ through relation
(\ref{LDI_definition}) is the following: According to
\cite{Skokos_et._al_GALI} it is possible to determine whether an
orbit is chaotic or lies on a $d$ -- dimensional torus by choosing
$m$ deviation vectors and computing the GALI$_m$ index. If GALI$_m
\approx\mbox{const.}$ for $m=2,3,\ldots,d$ and for $m>d$ decay by a
power law, the motion lies on a $d$ -- dimensional torus. If, on the
other hand, all GALI$_m$ indices decay exponentially the motion is
chaotic. Thus, to characterize orbits we often have to compute
GALI$_m$ indices for $m$ as high as $N$ or higher.

A serious limitation appears, of course, in the case of Hamiltonian
systems of large $N$, where GALI$_N(t)$ involves the computation of
$\begin{pmatrix}
                               2N \\
                               N \\
                             \end{pmatrix}=\frac{(2N)!}{(N!)^2}$
determinants at every time step. For example, in a Hamiltonian
system of $N=15$ dof, $\mathrm{GALI}_{15}(t)$ requires, for a given
orbit, the computation of $155117520$ determinants at every time
step while $\mathrm{LDI(t)=LDI_{15}}(t)$ requires only the
application of the SVD method for a $30\times 15$ matrix
$\mathcal{A}(t)$!

Clearly, at every point of the orbit $\vec{x}(t)$ the $2\leq
m\leq2N$ deviation vectors span a subspace of the $2N$ --
dimensional tangent space of the orbit, which is isomorphic to the
Euclidean $2N$ -- dimensional phase space of the Hamiltonian system
(\ref{2}). Thus, if $k$ of the $m$ singular values
$w_k(t),k=1,\ldots,m$ are equal to zero, then $k$ columns of matrix
$\mathcal{A}(t)$ of deviation vectors are linearly dependent with
the remaining ones and the subspace spanned by the column vectors of
matrix $\mathcal{A}(t)$ is $d(=m-k)$ -- dimensional.

From a more geometrical point of view, let us note that the $m$
variational equations (\ref{3}) combined with the equations of
motion (\ref{2}) describe the evolution of an initial $m$ --
dimensional unit ball into an $m$ (or less) dimensional ellipsoid in
the tangent space of the Hamiltonian flow. Now, the deviation
vectors $\vec{\upsilon_i}(t)$ forming the columns of
$\mathcal{A}(t)$ do not necessarily coincide with the ellipsoid's
principal axes. On the other hand, in the case of a chaotic orbit,
every generically chosen initial deviation vector has a component in
the direction of the maximum (positive) Lyapunov exponent, so that
all initial tangent vectors in the long run, will be aligned with
the longest principal axis of the ellipsoid. The key idea behind the
LDI method is to take advantage of this fact to overcome the costly
calculation of the many determinants arising in the GALI$_m$ method
and characterize a reference orbit as chaotic or not, via the trends
of the stretching and shrinking of the $m$ principal axes of the
ellipsoid.

Thus, LDI solves the problem of orbit characterization by finding
new orthogonal axes for the ellipsoid at every time step and taking
advantage of the SVD method. Since the matrix $V$ in (\ref{3}) is
orthogonal, we have $V^{\top}=V^{-1}$, so that equation
(\ref{SVD_based_theorem}) gives:
\begin{equation}\label{7}
\mathcal{A}_{2N\times m}\cdot V_{m\times m}=U_{2N\times m}\cdot
W_{m\times m}
\end{equation}
at every time step. Geometrically, Eq. (\ref{7}) implies that the
image formed by the column vectors of matrix $V$ is equal to an
ellipsoid whose $i^{\mbox{th}}$ principal axis direction in the
tangent space of the reference orbit is given by:
\begin{equation}
w_i\cdot \textbf{$u_i$}
\end{equation}
where $w_i$ are the singular values of matrix $\mathcal{A}(t)$ and
$\textbf{$u_i$}$ is the $i^{\mbox{th}}$ column of matrix $U(t)$.
This is, in fact, the content of a famous theorem stating that:
\begin{theorem}(\cite{Alligood})
Let $\mathcal{A}$ be a $2N\times m$ matrix, and let $U$ and $W$ be
matrices resulting from the SVD of $\mathcal{A}$. Then, the columns
of $\mathcal{A}$ span an ellipsoid whose $i^{\mbox{th}}$ principal
axis is $w_i\cdot \textbf{$u_i$}$, where
$W=diag(w_1,w_2,\ldots,w_m)$ (singular values) and
$\{\textbf{$u_i$}\}_{i=1}^{m}$ are the columns of $U$.
\end{theorem}

According to this theorem, the principal axes of the ellipsoid
created by the time evolution of equation (\ref{3}) in the tangent
space of the reference orbit $\vec{x}(t)$ at every time $t$, are
stretched or shrunk, according to the singular values of $w_i>1$
or $w_i<1$ respectively for $i=1,\ldots,m$.

If it so happens that $k$ of the singular values $w_i=0$ as $t$
grows, then the corresponding principal axes of the ellipsoid vanish
and the ellipsoid is less than $m$ -- dimensional in the tangent
space of the reference orbit because the corresponding deviation
vectors of matrix $\mathcal{A}$ have become linearly dependent.

Thus, two distinct cases exist depending on whether the reference
orbit $\vec{x}(t)$ is chaotic or ordered:
\begin{enumerate}
\item{If the orbit is chaotic, the $m$ deviation vectors become
linearly dependent so that $\mathrm{GALI_m}(t)\rightarrow 0$
exponentially \cite{Skokos_et._al_GALI}. Consequently, at least
one of the singular values $w_i(t),i=2,\ldots,m$ becomes zero and
$\mathrm{LDI}_m(t)=\prod_{j=1}^m w_j(t)\rightarrow 0$ (also
$\mathrm{LDI}(t)\rightarrow 0$) for all $m\geq i$}.

\item{If the orbit is ordered (i.e. quasiperiodic) lying on a $d$ -- dimensional torus,
there is no reason \cite{Skokos_et._al_GALI,Skokos_et._al_PTPS} for
the $m$ deviation vectors to become linearly dependent, as long as
$m\leq d$. No principal axis of the ellipsoid is eliminated, since
all singular values $w_i,\;i=1,\ldots,m$ are nonzero and
$\mathrm{LDI_m}(t)$ fluctuates around nonzero positive values. On
the other hand, for $m\geq d$, the singular values
$w_i,\;i=d+1,\ldots,m$ tend to zero following a power law
\cite{Skokos_et._al_GALI}, since $m-d$ deviations will eventually
become linearly dependent with those spanning the $d$ -- dimensional
tangent space of the torus
\cite{Skokos_et._al_GALI,Skokos_et._al_PTPS}}.
\end{enumerate}
In the remainder of the paper, we apply the LDI indices and
numerically demonstrate that:
\begin{equation}\label{LDI_m}
\mathrm{LDI}_m=\mathrm{GALI}_m,\quad m=2,\ldots,2N
\end{equation}
for the same choice of $m$ initially linearly independent deviation
vectors $\vec{\upsilon_i}(0),\;i=1,\ldots,m$. In particular, we
present evidence that supports the validity of relation
(\ref{LDI_m}) and exploit it to identify rapidly and reliably
ordered and chaotic orbits in a $1$ -- dimensional, $N$ degree of
freedom Fermi -- Pasta -- Ulam lattice under fixed and periodic
boundary conditions
\cite{Antonopoulos_et_al_IJBC,Antonopoulos_et._al_PRE}. We propose
that the validity of (\ref{LDI_m}) is due to the fact that both
quantities measure the volume of the same ellipsoid, the difference
being that, in the case of the LDI, the principal axes of the
ellipsoid are orthogonal. As we have not proved it, however, this is
a point to which we intend to return in a future publication.
\section{Application to the FPU Hamiltonian
System}\label{FPU_example}

In this section, we apply the LDI method to the case of a
multidimensional Hamiltonian system. Our aim is the comparison of
its performance and effectiveness in distinguishing between ordered
and chaotic behavior compared with Lyapunov exponents as well as the
SALI and GALI methods.

We shall use the $N$ dof Hamiltonian system of the $1$D lattice of
the Fermi -- Pasta -- Ulam (FPU) $\beta$ -- model. The system is
described by a Hamiltonian function containing quadratic and quartic
nearest neighbor interactions:
\begin{equation}\label{FPU_Hamiltonian}
H_N=\frac{1}{2}\sum_{j=1}^{N}\dot{x}_{j}^{2}+\sum_{j=0}^{N}\biggl
(\frac{1}{2}(x_{j+1}-x_{j})^2+\frac{1}{4}\beta(x_{j+1}-x_{j})^4\biggr)=E
\end{equation}
where $x_{j}$ is the displacement of the $j^{\mbox{th}}$ particle
from its equilibrium position, $\dot{x}_{j}$ is the corresponding
conjugate momentum, $\beta$ is a positive real constant and $E$ is
the constant energy of the system.

\begin{figure}[h!]
\begin{center}
\includegraphics[scale=0.85]{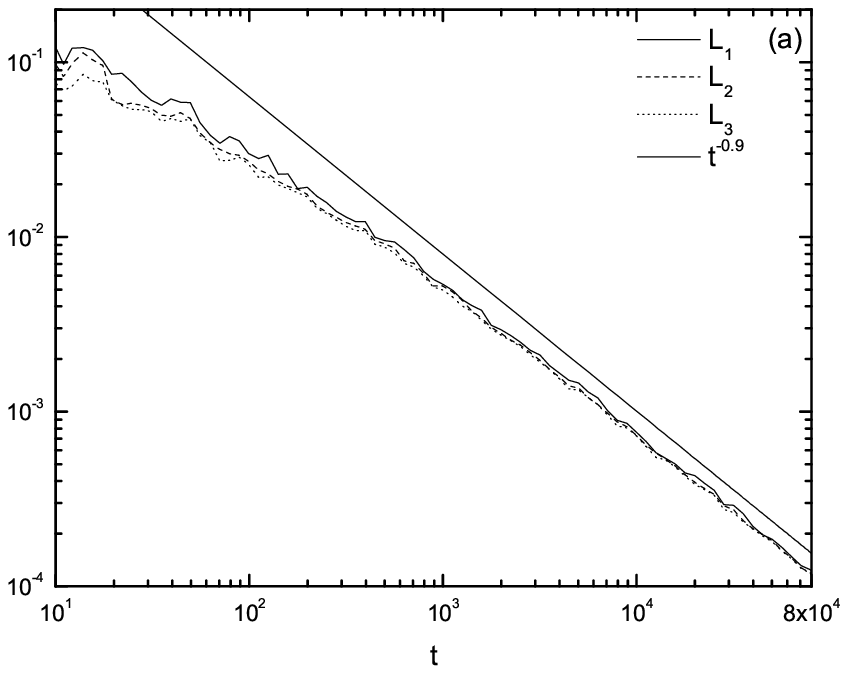}
\includegraphics[scale=0.85]{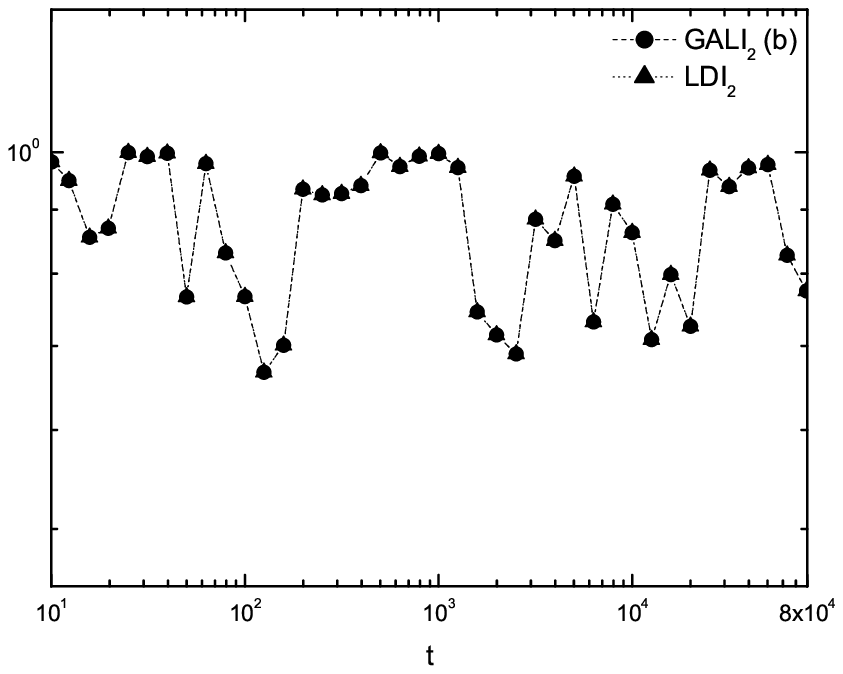}
\includegraphics[scale=0.85]{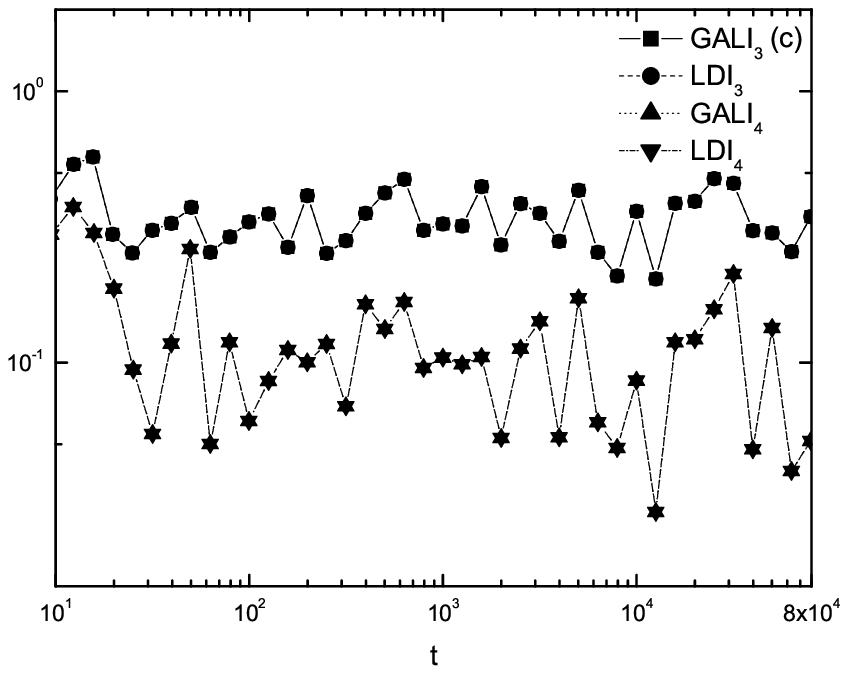}
\includegraphics[scale=0.841]{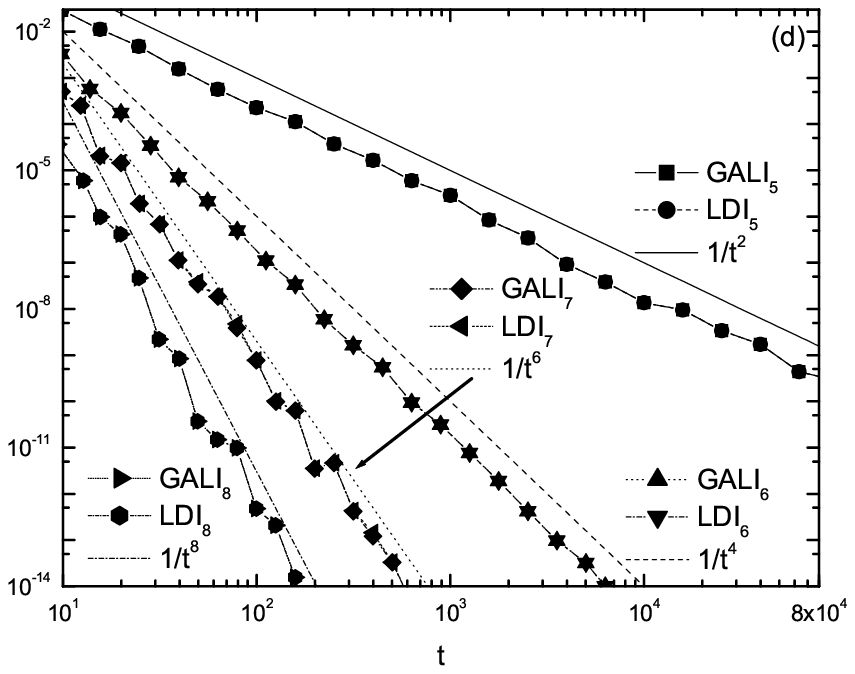}
\end{center}
\caption{The case of an ordered orbit:(a) The time evolution of the
three maximal positive \Ly. (b) The time evolution of \GA$_2$ and
LDI$_2$. (c) The time evolution of \GA$_3$, LDI$_3$ and \GA$_4$,
LDI$_4$. (d) The time evolution of \GA$_m$, LDI$_m$ with
$m=5,\ldots,8$ and the corresponding slopes of the fall to zero. In
all panels we have chosen a neighboring orbit at a distance
$\approx2.1$ from the OPM of the FPU Hamiltonian
(\ref{FPU_Hamiltonian}) with periodic boundary conditions for $N=4$
and $E=2$. All axes are logarithmic.} \label{fig_2}
\end{figure}

We start by focusing on an ordered case choosing a neighboring orbit
of the stable out of phase mode (OPM)
\cite{Budinsky,Poggi,Antonopoulos_et_al_IJBC}, which is a simple
periodic orbit of the FPU Hamiltonian (\ref{FPU_Hamiltonian}). This
solution exists for every $N$, for fixed as well as periodic
boundary condition (PBC):
\begin{equation}\label{FPU_periodic_boundary_conditions_OPM}
x_{N+1}(t)=x_{1}(t),\;\forall t
\end{equation}
and is given by:
\begin{equation}\label{FPU_non_lin_mode_periodic_boundary_conditions_OPM}
x_{j}(t)=-x_{j+1}(t),\;\dot{x}_j(t)=0,\;j=1,\ldots,N,\forall t.
\end{equation}

In \cite{Budinsky,Antonopoulos_et_al_IJBC} the stability properties
of the OPM mode with periodic boundary conditions were determined
using Floquet theory and monodromy matrix analysis and the energy
range $0\leq E(N)\leq E_{c}^{\mathrm{OPM}}(N)$ over which it is
linearly stable was studied in detail.

It is known that for $N=4$ and $\beta=1$, the solution
(\ref{FPU_non_lin_mode_periodic_boundary_conditions_OPM}) with
periodic boundary condition
(\ref{FPU_non_lin_mode_periodic_boundary_conditions_OPM}) is
destabilized for the first time at the critical energy
$E_c^{\mathrm{OPM}}\approx4.51$. Below this critical energy, the OPM
is linearly stable and is surrounded by a sizable island of
stability. By contrast, for $E>E_c^{\mathrm{OPM}}$, the OPM is
linearly unstable with no island of stability around it.

In Fig. \ref{fig_2}(a), we have calculated the three maximal
Lyapunov exponents of a neighboring orbit located at distance
$\approx2.1$ away from the OPM at $E=2<E_c^{\mathrm{OPM}}$. At this
energy, the OPM is linearly stable and thus all Lyapunov Exponents
tend to zero following a simple power law. Next, in Fig.
\ref{fig_2}(b), we compute \GA$_2$ and LDI$_2$ for a final
integration time $t=8\times10^4$ and observe that \GA$_2$ and
LDI$_2$ practically coincide fluctuating around non zero values
indicating the ordered nature of the orbit. \GA$_2$ needs $558$
seconds of computation time while LDI$_2$ takes about $912$ seconds
in a Pentium 4 3.2GHz computer.

In Fig. \ref{fig_2}(c), we compute \GA$_3$, LDI$_3$ and \GA$_4$,
LDI$_4$ for the same energy and initial condition. We see once more
that \GA$_3$, LDI$_3$ and \GA$_4$, LDI$_4$ coincide fluctuating
around non zero values. The \GA$_3$ computation now takes about
$1044$ seconds, LDI$_3$ about $838$ seconds, \GA$_4$ needs $898$
seconds and LDI$_4$ $753$ seconds.

Finally, in Fig. \ref{fig_2}(d), we present \GA$_m$, LDI$_m$ with
$m=5,\ldots,8$ as a function of time for the same energy and initial
condition. We observe again that \GA$_m$ and LDI$_m$ with
$m=5,\ldots,8$, have the same values and tend to zero following a
power law of the form $t^{-2(k-N)}$. All these results are in
accordance with the formulae reported in \cite{Skokos_et._al_GALI}
and suggest that the torus on which the orbit lies is 4 --
dimensional, as expected from the fact that the number of dof of the
system is $N=4$.

In \cite{Antonopoulos_et_al_IJBC} we also studied the stability
properties of a different simple periodic orbit of FPU called the
SPO1 mode with fixed boundary conditions (FBC). Using monodromy
matrix analysis we found that for $N=5$ and $\beta=1.04$, the SPO1
mode with FBC is destabilized for the first time at the critical
energy $E_c^{\mathrm{SPO1}}\approx6.4932$.

Thus, in order to study a chaotic case where things are different,
we choose initial condition at distance of
$\approx1.27\times10^{-4}$ from the SPO1 orbit at the energy $E=11$,
where it is unstable.

\begin{figure}[h!]
\begin{center}
\includegraphics[scale=0.85]{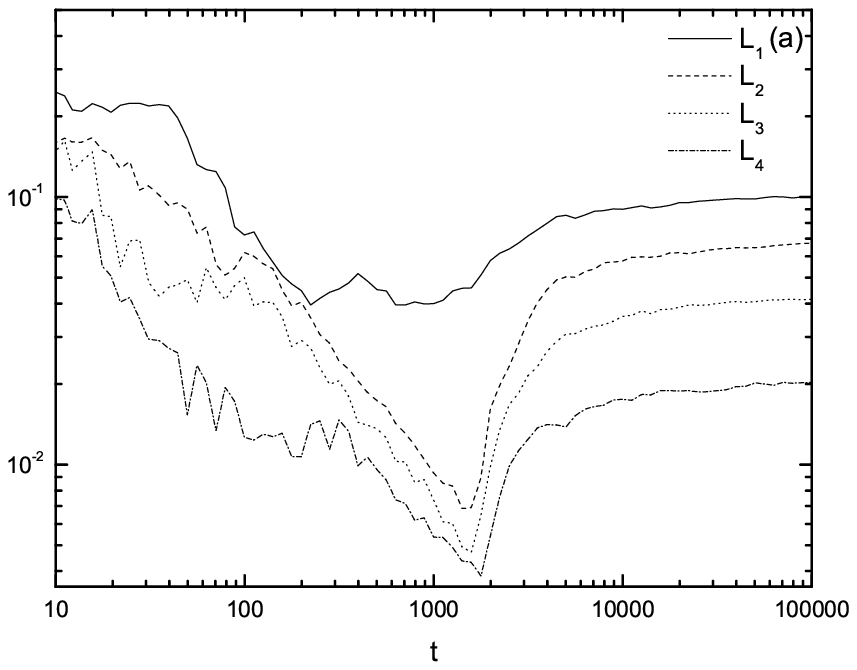}
\includegraphics[scale=0.85]{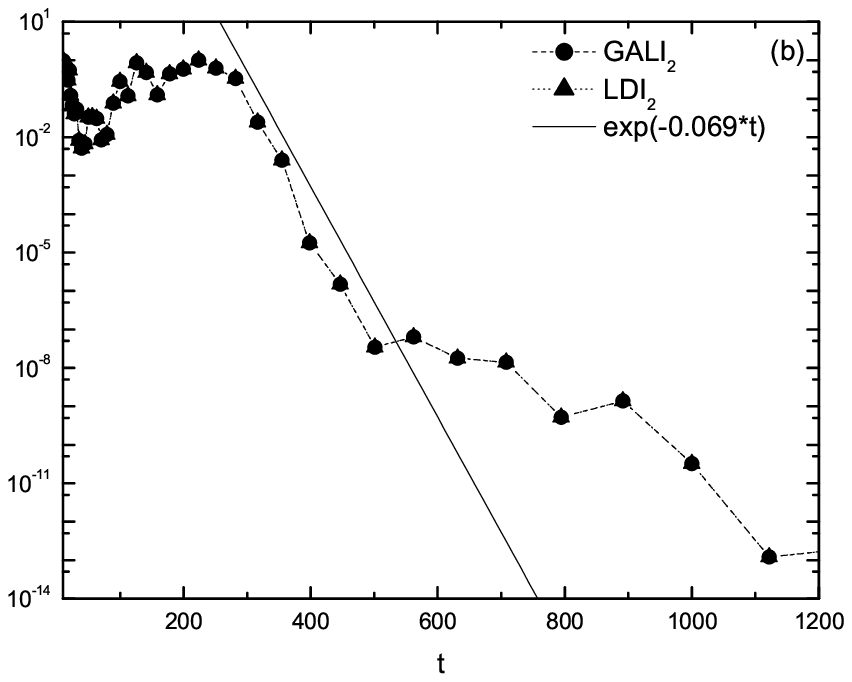}
\includegraphics[scale=0.85]{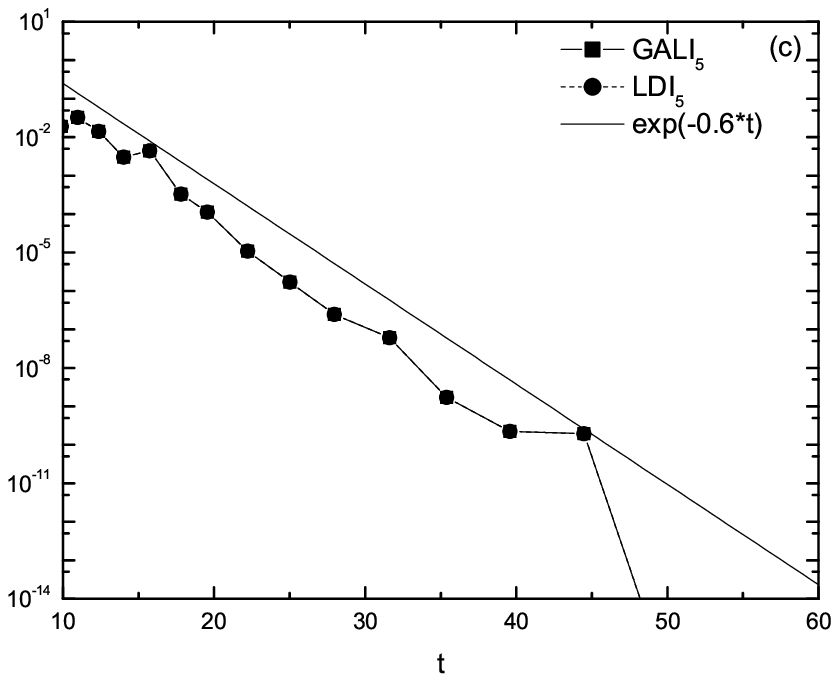}
\end{center}
\caption{The case of a chaotic orbit:(a) The time evolution of the
four maximal \Ly. (b) The time evolution of \GA$_2$, LDI$_2$ follows
the approximate formula $e^{-(\sigma_1-\sigma_2)t}$ where
$\sigma_1\approx0.124$ (solid straight line) and
$\sigma_2\approx0.056$ for $t=71$. (c) The time evolution of
\GA$_5$, LDI$_5$ follows the approximate formula $\propto
e^{-[(\sigma_1-\sigma_2)+(\sigma_1-\sigma_3)+(\sigma_1-\sigma_4)+(\sigma_1-\sigma_5)]t}\approx
e^{-0.069t}$ (solid straight line) where $\sigma_1\approx0.197$,
$\sigma_2\approx0.095$, $\sigma_3\approx0.047$,
$\sigma_4\approx0.026$ and $\sigma_5\approx0.022$ for $t\approx44$.
We have used, in all figures, the same orbit of a distance of
$1.27\times10^{-4}$ from the SPO1 of the FPU Hamiltonian system
(\ref{FPU_Hamiltonian}) with fixed boundary conditions for $N=5$ and
$E=11$.} \label{fig_3}
\end{figure}

In Fig. \ref{fig_3}(a), we calculate \Ly~of the above mentioned
orbit and find that the four maximal \Ly~tend to positive values.
This is strong evidence that the nature of the orbit is chaotic.
Next, in Fig. \ref{fig_3}(b) we calculate \GA$_2$ and LDI$_2$ up to
$t=1200$. We see the indices again coincide and tend to zero as
$\propto e^{-(\sigma_1-\sigma_2)t}$ (solid straight line), as
predicted by our theory \cite{Skokos_et._al_JPA,Skokos_et._al_GALI}.
In this figure, we find $\sigma_1\approx0.124$ and
$\sigma_2\approx0.056$ for time $t=71$. The corresponding CPU time
required for the calculation of all indices does not differ
significantly, as they become quite small in magnitude, rather
quickly.

Nevertheless, LDI$_2$ requires less CPU time than \GA$_2$. In Fig.
\ref{fig_3}(c), we calculate \GA$_5$ and LDI$_5$ for the same energy
and initial condition as in the previous panels. We observe now that
\GA$_5$ and LDI$_5$ coincide falling to zero as \GA$_5\propto
e^{-[(\sigma_1-\sigma_2)+(\sigma_1-\sigma_3)+(\sigma_1-\sigma_4)+(\sigma_1-\sigma_5)]t}$
(solid straight line) where $\sigma_1\approx0.197$,
$\sigma_2\approx0.095$, $\sigma_3\approx0.047$,
$\sigma_4\approx0.026$ and $\sigma_5\approx0.022$ for $t\approx44$.
Clearly, \GA$_5$ and LDI$_5$ distinguish the chaotic character of
the orbit faster than \GA$_2$ or LDI$_2$. This is so, because
\GA$_2$ or LDI$_2$ reaches the threshold $10^{-8}$
\cite{Skokos_et._al_PTPS,Skokos_et._al_JPA,Skokos_et._al_GALI} for
$t\approx750$ while \GA$_5$ and LDI$_5$ for $t\approx35$! The CPU
times required for the calculation of \GA$_5$ and LDI$_5$ up to
$t=80$ are approximately $1.5$ seconds each.

Thus, we conclude from these results that the LDI method performs at
least as well as the GALI, predicting correctly the ordered or
chaotic nature of orbits in Hamiltonian systems for low dimensions,
i.e. at 2, 4 and 5 degrees of freedom. However, in higher
dimensional cases, GALI indices become very impractical as they
demand the computations of millions of determinants at every time
step making the LDI method much more useful.

In order to show the advantages of the LDI method concerning the CPU
time, we repeat the above analysis for the same Hamiltonian system
(\ref{FPU_Hamiltonian}), but now for $N=15$ and energy $E=2$, and
for an initial condition very close to the unstable SPO1
\cite{Antonopoulos_et_al_IJBC}.

\begin{figure}[h!]
\begin{center}
\includegraphics[scale=0.845]{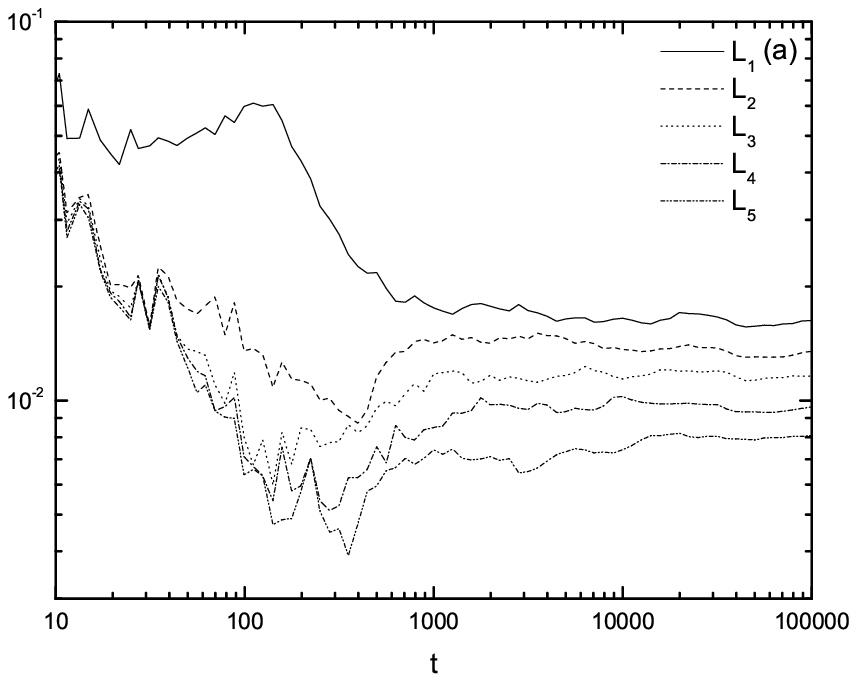}
\includegraphics[scale=0.845]{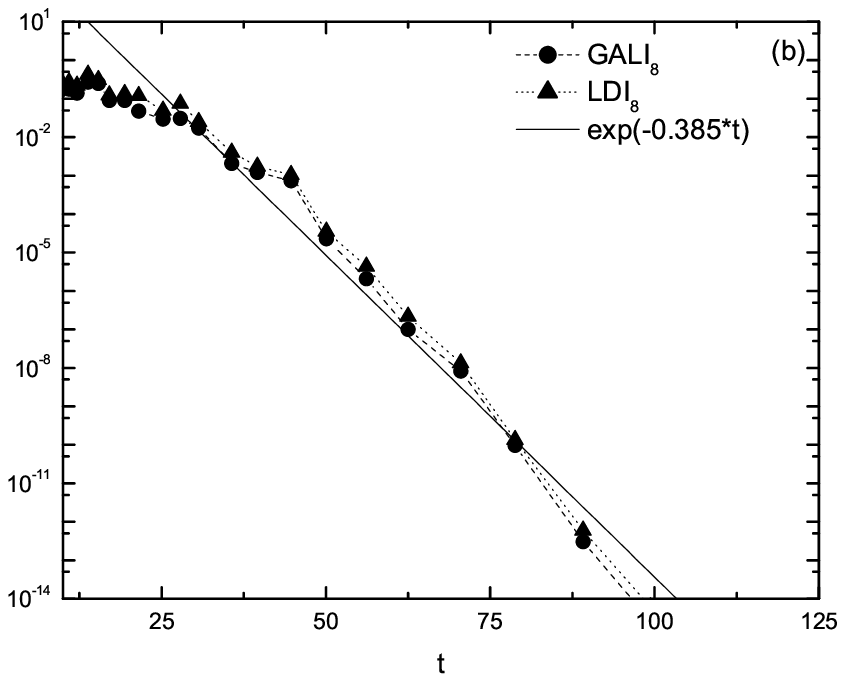}
\end{center}
\caption{(a) The time evolution of the five positive \Ly. (b) The
time evolution of the \GA$_8$, LDI$_8$ follows the approximate
formula $\propto
e^{-[(\sigma_1-\sigma_2)+(\sigma_1-\sigma_3)+\ldots+(\sigma_1-\sigma_8)]t}\approx
e^{-0.385t}$ (solid straight line) where $\sigma_1\approx0.061$,
$\sigma_2\approx0.011$, $\sigma_3\approx0.006$,
$\sigma_4\approx0.005$, $\sigma_5\approx0.005$,
$\sigma_6\approx0.004$, $\sigma_7\approx0.004$ and
$\sigma_8\approx0.004$ for time $t\approx141$. In all panels we have
used initial conditions at a distance of $9\times10^{-5}$ from the
SPO1 orbit of Hamiltonian system (\ref{FPU_Hamiltonian}) with fixed
boundary conditions, $N=15$ and $E=2$.} \label{fig_4_LDI}
\end{figure}

In \cite{Antonopoulos_et_al_IJBC} it has also been shown that for
$N=15$ and $\beta=1.04$, the SPO1 with fixed boundary conditions
destabilizes at the critical energy $E_c\approx1.55$. Thus, for
energies smaller than $E_c$, SPO1 is linearly stable, while for
$E>E_c$ it is unstable and is surrounded by a chaotic region.

In Fig. \ref{fig_4_LDI}(a) we depict the time evolution of the five
maximal \Ly~which converge to positive values for high enough $t$
suggesting that the neighboring orbit is chaotic. In the second
panel of the same figure we present the evolution of \GA$_8$ and
LDI$_8$ together with the approximate exponential law. We remark
once more that the values of the corresponding indices coincide
until they become numerically zero. More interestingly, the CPU time
required for the calculation of \GA$_8$ up to $t\approx100$ is about
$186$ seconds while for the LDI$_8$ it takes only one second! This
difference is very important, showing why LDI is preferable compared
to the corresponding \GA~index in Hamiltonian systems of many
degrees of freedom.

\section{Conclusions}\label{conclusions}

In this paper we have introduced a new method for distinguishing
quickly and reliably between ordered and chaotic orbits of
multidimensional Hamiltonian systems and argued about its validity
justifying it in the ordered and chaotic case. It is based on the
recently introduced theory of the Generalized Alignment Indices
(GALI). Following this theory, the key point in the distinction
between order and chaos is the linear dependence (or independence)
of deviation vectors from a reference orbit. Consequently, the
method of LDI takes advantage of this property and analyzes $m$
deviation vectors using Singular Value Decomposition to decide
whether the reference orbit is chaotic or ordered. If the orbit
under consideration is chaotic then the deviation vectors are
aligned with the direction of the maximal Lyapunov exponent and thus
become linearly dependent. On the other hand, if the reference orbit
is ordered then there is no unstable direction and $m=1,2,...,d\leq
N$ deviation vectors are linearly independent. As a consequence, the
LDI of order $m$ (LDI$_m$) becomes either zero if the reference
orbit is chaotic or it fluctuates around non zero values if the
orbit is ordered if $m\leq d$.

After introducing the new method, we presented strong numerical
evidence about its validity and efficiency in the interesting case
of multidimensional Hamiltonian systems. One first main result is
that GALI$_m$ and LDI$_m$ coincide numerically for the same $m$
number of deviation vectors and for the same reference orbit.
Moreover, it follows that it is preferable to use the LDI method
rather than the equivalent GALI method especially in the
multidimensional case of Hamiltonian systems, since the LDI needs
considerably less CPU time than the corresponding GALI method for
the same number of deviation vectors.


\section{Acknowledgements}

This work was partially supported by the European Social Fund (ESF),
Operational Program for Educational and Vocational Training II
(EPEAEK II) and particularly the Program PYTHAGORAS II. We thank Dr.
Charalambos Skokos and Miss Eleni Christodoulidi for very fruitful
discussions on the comparison between the GALI and LDI indices.

\bibliography{bibliography}

\end{document}